\documentclass{article}
\usepackage[utf8]{inputenc}
\usepackage[T1]{fontenc}
\usepackage{hyperref}
\usepackage{natbib}

\usepackage{blindtext}
\usepackage{subfig}
\usepackage[toc,page]{appendix}
\usepackage{booktabs}
\usepackage{float}
\usepackage{natbib}
\usepackage{csquotes}
\usepackage{graphicx}
\usepackage{adjustbox}

\title{Leveraging Computer Vision Application in Visual Arts:
A Case Study on the Use of Residual Neural Network to
Classify and Analyze Baroque Paintings}
\author{\textbf{Daniel Kvak}\footnote{Corresponding author: kvak@mail.muni.cz}\\
	Faculty of Arts\\
	Masaryk University\\
	Brno, Czech Republic\\\\
	ORCID: 0000-0001-7808-7773}

\begin{document}

\maketitle
\begin{abstract}
    With the increasing availability of large digitized fine art collections, automated analysis and classification of paintings is becoming an interesting area of research. However, due to domain specificity, implicit subjectivity, and pervasive nuances that vaguely separate art movements, analyzing art using machine learning techniques poses significant challenges. Residual networks, or variants thereof, are one the most popular tools for image classification tasks, which can extract relevant features for well-defined classes. In this case study, we focus on the classification of a selected painting 'Portrait of the Painter Charles Bruni' by Johann Kupetzky and the analysis of the performance of the proposed classifier. We show that the features extracted during residual network training can be useful for image retrieval within search systems in online art collections.
\end{abstract}


\newpage
\section{Introduction}
Image classification is one of the most widely used computer vision tasks. \citep{lu2007survey} In the recent past, deep learning has been very successful in various visual tasks, such as agent-based simulation of autonomous vehicles \citep{schwarting2018planning} or computer-aided detection / diagnosis in the healthcare segment. \citep{doi2007computer} The extensive digitization that has occurred in the last two decades \citep{aydougan2019reflections} has led to the question of whether the curation segment can also be automated using machine methods. The conversion of information from physical works of art into digital image format plays a key role in the opening of new research challenges in the interdisciplinary field of computer vision, machine learning, and art history. \citep{cetinic2018fine, tan2016ceci, saleh2015large}

Different convolutional neural network (CNN) architectures have been proven to work well for image recognition and classification tasks. The basic idea is that neurons in the visual cortex process images into increasingly complex shapes. \citep{lindsay2021convolutional} The image is first segmented at edge boundaries using a light / dark interface, then merged into simple shapes, and finally merged into recognizable complex features in subsequent layers. \citep{albawi2017understanding} Individual class labels may be based on some low-level features such as color, texture, or shape, but are most often based on higher-level features such as semantic description, activity, or artistic style. \citep{o2015introduction} CNN tries to mimic this idea using several layers of artificial neurons. The standard architecture includes several convolutional layers that segment the image into small chunks that can be easily processed. \citep{albawi2017understanding}

\section{Proposed Method}
The use of machine learning for automatic classification of fine art collections has received little attention in the literature so far. \citep{arora2012towards, rodriguez2018classification} In recent years, libraries, museums, galleries, and art centers have been digitizing their collections to promote public interest in the arts and facilitate access to masterpieces from the comfort of home, a trend that has been further reinforced by the ongoing COVID-19 pandemic. \citep{habsary2021digitalization} These activities create a demand for automated analysis and classification of digitized art. \citep{khoronko2021museum} In this paper, we propose a novel approach to using CNN output to classify visual artwork. Using CNN pre-trained on ImageNet,\footnote{ImageNet is a large-scale visual database designed for use in image classification and object recognition research. The project includes more than 14 million images that have been manually annotated to indicate what objects are shown. ImageNet features more than 20,000 categories, with a typical category such as "balloon" or "strawberry" consisting of several hundred images} we consider feature maps computed at the level of several different layers before fully connected layers and compare the perception of artificial intelligence with the analysis of art historians and curators. We show that the extracted features are effective for classifying artists and styles and provide a detailed visualization and discussion of the suitability and effectiveness of the different layers.

\subsection{Transfer Learning}
In transfer learning, a neural network is first trained on a generic dataset (e.g. ImageNet visual database), and the features learned from the initial task are transferred to a new network that is fine-tuned for a specific task. \citep{weiss2016survey} Deploying pre-trained models on similar data has shown solid results in image classification-related tasks. \citep{weiss2016survey, zhuang2020comprehensive} Several organizations have created models such as VGG \citep{sengupta2019going}, Inception \citep{szegedy2016rethinking}, or ResNet \citep{he2016deep} that would take weeks to train on user-accessible hardware. Pre-trained networks can be downloaded and easily fine-tuned to result in lower generalization error while using less computational effort.

\subsection{ResNet50V2 Model Architecture}
As deep learning evolves, the structure of neural networks deepens; while this helps the network to perform more complex feature extraction, it can also introduce the problem of vanishing or exploding gradients. \citep{joshi2019issues} This can lead to the following drawbacks: (1) Long training time with the convergence of the network becomes very difficult or even non-convergent. (2) The network performance gradually becomes saturated and even starts to decline. \citep{joshi2019issues, kim2016accurate}

\begin{figure}[!h]
\centering
\includegraphics[width=1\textwidth]{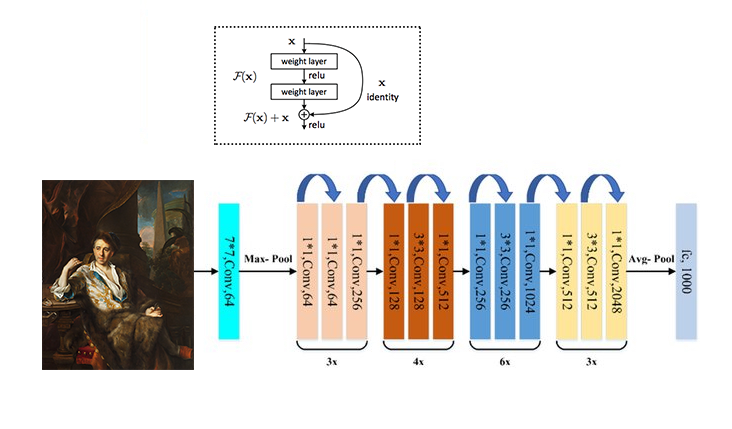}
\caption{\label{fig:resnet}\textbf\scriptsize{Proposed architecture of ResNet50V2 model.}}
\end{figure}

Instead of designing own architecture, we leverage insights from a set of existing convolutional neural network architectures that show excellent performance in solving a variety of classification tasks. Residual networks (ResNets) are a unique type of deep convolutional networks whose basic idea is to skip blocks of convolutional layers by using shortcut connections. \citep{he2016deep} In this case study, we use a variant of residual neural networks called ResNet50V2 (shown in \autoref{fig:resnet}). \citep{yamazaki2019yet}

The basic building blocks follow two simple rules: (i) for the same output feature map size, the layers have the same number of filters; and (ii) if the feature map size is halved, the number of filters is doubled. \citep{he2016deep} The down-sampling is performed by convolutional layers that have a stride of 2 and batch normalization is performed right after each convolution operation and before ReLU activation. \citep{he2016deep} When the input and output are of the same dimensions, the identity shortcut is used. When dimensions increase, the projection shortcut is used to match dimensions through 1 × 1 convolutions. In both cases, when the shortcuts go across feature maps of two sizes, they are performed with a stride of 2. \citep{he2016deep} The network output ends with a 1,000 fully-connected layer and softmax function. The total number of weighted layers is 50, with 25.6 million trainable parameters. \citep{yamazaki2019yet}

ResNets provide a trade-off between performance and number of parameters. \citep{huang2017speed} The weights used in the proposed model have been pre-trained using the ImageNet database. \citep{yamazaki2019yet} Our hypothesis is that despite the obvious discrepancy between images embedded in ImageNet databbase and fine art collections, ResNet-50V2 comprehensively pre-trained on the ImageNet may still be transferred to perform style classification tasks more effective. 

\begin{table}
\begin{adjustbox}{width=1\textwidth}
\begin{tabular}{llllll}
\toprule
\textbf{\citep{he2016deep}} & \textbf{Layers} & \textbf{Depth} & \textbf{Parametres} & \textbf{Input matrix} & \textbf{Output activation} \\ \midrule
ResNet50V2   & 50              & 103            & 25.6M                & 224, 224, 3                      & softmax                    \\ \bottomrule
\end{tabular}
\end{adjustbox}
\caption{\label{tab:table-name}Architecture of the proposed model}
\end{table}

\subsection{Feature Extraction}
\begin{figure}[H]
\centering
\includegraphics[width=1\textwidth]{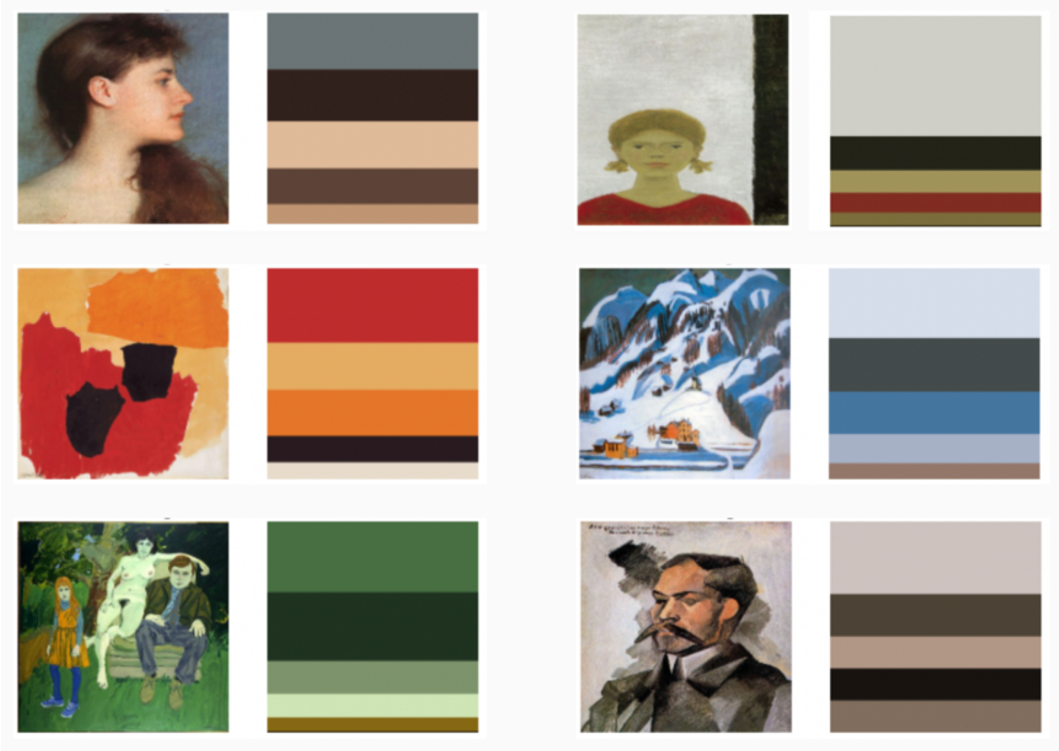}
\caption{\label{fig:colors}\textbf\scriptsize{Dominant color palette of represented examples.}}
\end{figure}

The classification of painting styles is usually done by art historians and curators based on the relationship between subjective attributes, physical characteristics, as shown in \autoref{fig:colors} (light, lines, colors, textures, shapes, space, etc.), and appropriate historical periods. \citep{zhao2021compare} In many cases, however, significant stylistic differences can be observed, such as seamless transitions between artistic movements over time, stylistic differences between paintings by the same artist, unique personal characteristics that do not belong to any one style or artistic period, the influence of one artist on others, and varying interpretations of abstract and surreal elements. Therefore, these variables make it difficult to apply classification methods in the visual arts. \citep{sandoval2019two}
Feature extraction is part of the dimensionality reduction process, where the initial raw data set is split and transformed into more manageable groups. \citep{chen2016deep} Feature extraction aims to reduce the number of features in the dataset by creating new features from existing ones. This new reduced feature set should be able to summarize most of the information contained in the original feature set. \citep{wiatowski2017mathematical}

Image processing is one of the domains where feature extraction finds wide application. \citep{nixon2019feature} Feature extraction in CNN uses many techniques that include methods to detect low-level features such as colors, brightness, edges, or textures in order to process them. \citep{chen2016deep} The convolution layer consists of a set of digital filters that perform convolution operations on the input data. \citep{he2016deep} The convolutional layer serves as a dimensionality reduction layer and decides the threshold. During backpropagation, a number of parameters need to be adjusted, which in turn minimizes the coupling within the neural network architecture. \citep{he2016deep}

\begin{figure}[H]
\centering
\includegraphics[width=1\textwidth]{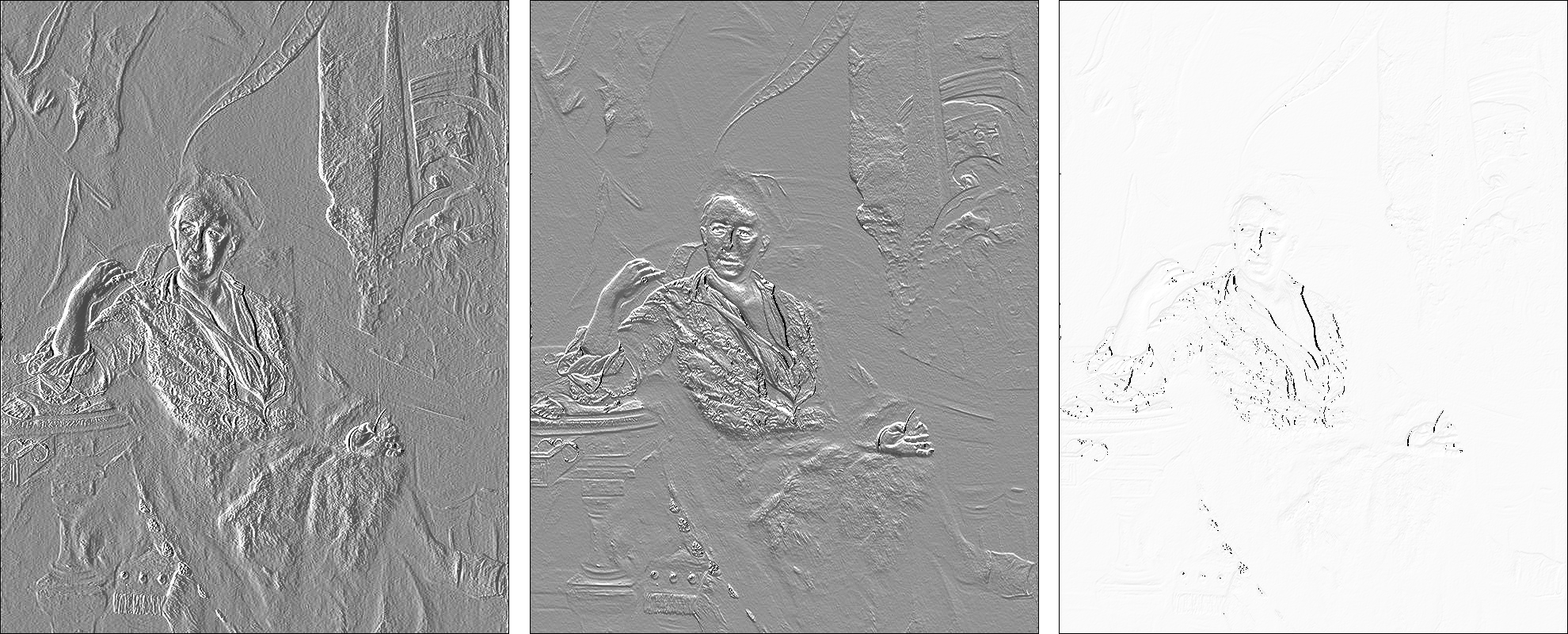}
\caption{\textbf\scriptsize{Principal Component Analysis of luminance gradient feature of ’Portrait of the Painter Charles Bruni’ by Johann Kupetzky.}}
\end{figure}

\section{Experiment}
\subsection{Dataset}
Since each class of styles contains paintings by different artists, training and classification is not simple. \citep{sandoval2019two} The ability to correctly classify artworks into narrowly defined domains is a difficult task even for curators of existing datasets. \citep{zhao2021compare} The achieved classification results are based on the freely available data set 'Art Snobs Data2' extracted from Kaggle. \citep{martin_2019} These paintings were collected from various sources and contain a large collection of art pieces from different eras, with 19 distinct styles, namely: Abstract Expressionism, Art Nouveau, Baroque, Cubism, Early Renaissance, Expressionism, High Renaissance, Impressionism, Mannerism, Naive Art, Neoclassicism, Northern Renaissance, Post-Impressionism, Realism, Rococo, Romanticism, Surrealism, Symbolism and Ukiyo-e. To perform style classification, we used 923 images from each class to train the classifiers and used the remaining 102 as the validation set, whereas according to Fig. 4, it is a perfectly balanced set. Thus, the dataset consists a total of 17,537 train images and 1,938 validation images.

\begin{figure}[H]
\centering
\includegraphics[width=0.8\textwidth]{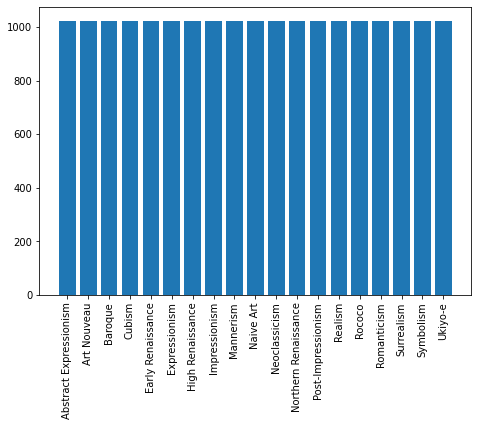}
\caption{\textbf\scriptsize{Distribution of train and validation examples for individual classes.}}
\end{figure}

\subsection{Data Augmentation}
Data augmentation increases the size of the input training data along with the regularization of the model, thus improving the generalization of the training model. \citep{shorten2019survey} It also helps to create new train examples by randomly applying different transformations to the available dataset to reflect the noisiness of real-world data. \citep{perez2017effectiveness} In our experiment, we used transformations that involve vertical flipping of training images, random rotations, modifications in lighting conditions, zoom, saturation, and JPEG encoding noise. The training data was augmented using five randomly selected variations; the extension of validation dataset was not investigated.

\subsection{Training}
The neural network learns using the backpropagation method. \citep{he2016deep} The weights of fully-connected layer of ResNet50V2 can be fine-tuned, i.e., this layer adapts by backpropagation, while the other layers of the network are invariant after pre-training on ImageNet database. \citep{yamazaki2019yet} Fine-tuning of the top layer in the ResNet50V2 is performed as it is not guaranteed that the mean and variance of these layers will be similar to the mean and variance of our dataset.

An F1 Score becomes a critical evaluation tool to determine False Positive and False Negative rates yielded through a discriminating threshold in a similar situation with unbalanced dataset samples. \citep{goutte2005probabilistic} The classification performance of our model for multi-class problem was evaluated for each component and the average classification performance of the model was calculated. \autoref{tab:testdata} includes the precision, recall, and F1 Score, calculated based on the following equations:

\begin{equation} \label{eqn2} 	 
Precision = \frac{TP}{TP+FP}
\end{equation}
\begin{equation} \label{eqn3} 
Recall = \frac{TP}{TP+FN}
\end{equation}
\begin{equation} \label{eqn4} 
F1 Score = \frac{2*Precision*Recall}{Precision+Recall} = \frac{2*TP}{2*TP+FP+FN}
\end{equation}
Based on the equations, the table below shows the Precision, Recall, F1 Score results for the multi-class classification:
\begin{table}[!h]
\centering
\begin{tabular}{@{}llll@{}}
\toprule
\textbf{Class}         & \textbf{Precision} & \textbf{Recall} & \textbf{F1 Score} \\ \midrule
Abstract Expressionism & 0.88               & 0.89            & 0.88              \\
Art Nouveau            & 0.73               & 0.66            & 0.69              \\
Baroque                & 0.63               & 0.45            & 0.53              \\
Cubism                 & 0.87               & 0.80            & 0.84              \\
Early Renaissance      & 0.83               & 0.66            & 0.74              \\
Expressionism          & 0.56               & 0.69            & 0.62              \\
High Renaissance       & 0.41               & 0.72            & 0.52              \\
Impressionism          & 0.86               & 0.55            & 0.67              \\
Mannerism              & 0.59               & 0.60            & 0.60              \\
Naive Art              & 0.86               & 0.73            & 0.79              \\
Neoclassicism          & 0.70               & 0.70            & 0.70              \\
Northern Renaissance   & 0.68               & 0.61            & 0.64              \\
Post-Impressionism     & 0.64               & 0.66            & 0.65              \\
Realism                & 0.47               & 0.63            & 0.54              \\
Rococo                 & 0.61               & 0.79            & 0.69              \\
Romanticism            & 0.74               & 0.32            & 0.45              \\
Surrealism             & 0.76               & 0.75            & 0.75              \\
Symbolism              & 0.65               & 0.69            & 0.67              \\
Ukiyo-e                & 0.88               & 0.95            & 0.91              \\ \bottomrule
\end{tabular}
\caption{\label{tab:testdata}ResNet50V2 performance on test data.}
\end{table}

\begin{figure}[H]
\centering
\includegraphics[width=1\textwidth]{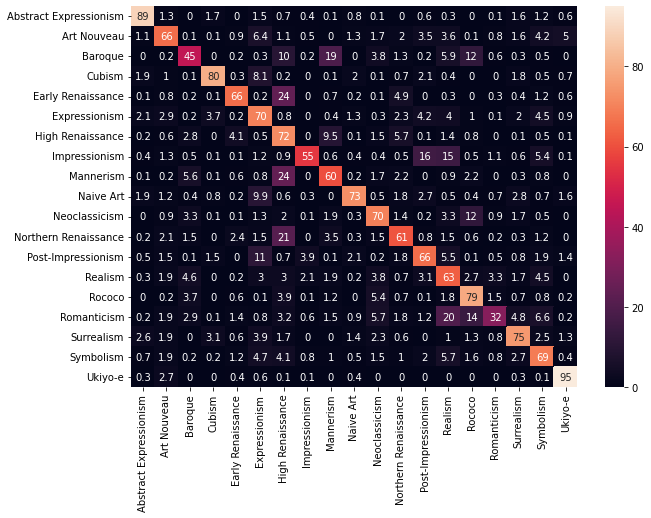}
\caption{\label{fig:conf}\textbf\scriptsize{Confusion matrix for multi-class classification.}}
\end{figure}

\section{Results}
The results obtained from the experiment can be analyzed and discussed from several perspectives. From the confusion matrix in \autoref{fig:conf}, we can observe the internal logic of the misclassified classes. The common visual features of the different styles explain the high misclassification rate (=lower model accuracy) between classes such as Northern Renaissance and High Renaissance \citep{nash2008northern} or Baroque and Rococo \citep{wittkower1999art} compared to other tasks, the lower classification of styles corresponds to the high overlap of visual features between classes and also to the large variety of content displayed in a single style. In contrast, the classes of the genre classification task are more uniform in content, and CNNs show a high ability to discriminate classes with radically different visual cues. \citep{hagtvedt2011turning}

This example of inaccurate classification, which can also be observed in the prediction for the analyzed work 'Portrait of the Painter Charles Bruni' by Johann Kupetzky, highlights the fact that style is not only associated with the mere visual characteristics and content of an artwork, but is often a subtly distinguishable and context-dependent concept, e.g. Baroque and Rococo suggested considerable confusion, which can be attributed to the fact that these styles are historically related. \citep{ruth1974classification}

\begin{table}[]
\centering
\begin{tabular}{@{}ll@{}}
\toprule
\textbf{Class} & \textbf{Confidence} \\ \midrule
Rococo         & 0.2942              \\ \midrule
Neoclassicism  & 0.2503              \\ \midrule
Baroque        & 0.1205              \\ \midrule
Realism        & 0.1201              \\ \midrule
Romanticism    & 0.1173              \\ \bottomrule
\end{tabular}
\caption{\label{tab:outcomes}Top 5 class predictions for 'Portrait of the Painter Charles Bruni' by Johann Kupetzky.}
\end{table}

\begin{figure}[H]
\centering
\includegraphics[width=1\textwidth]{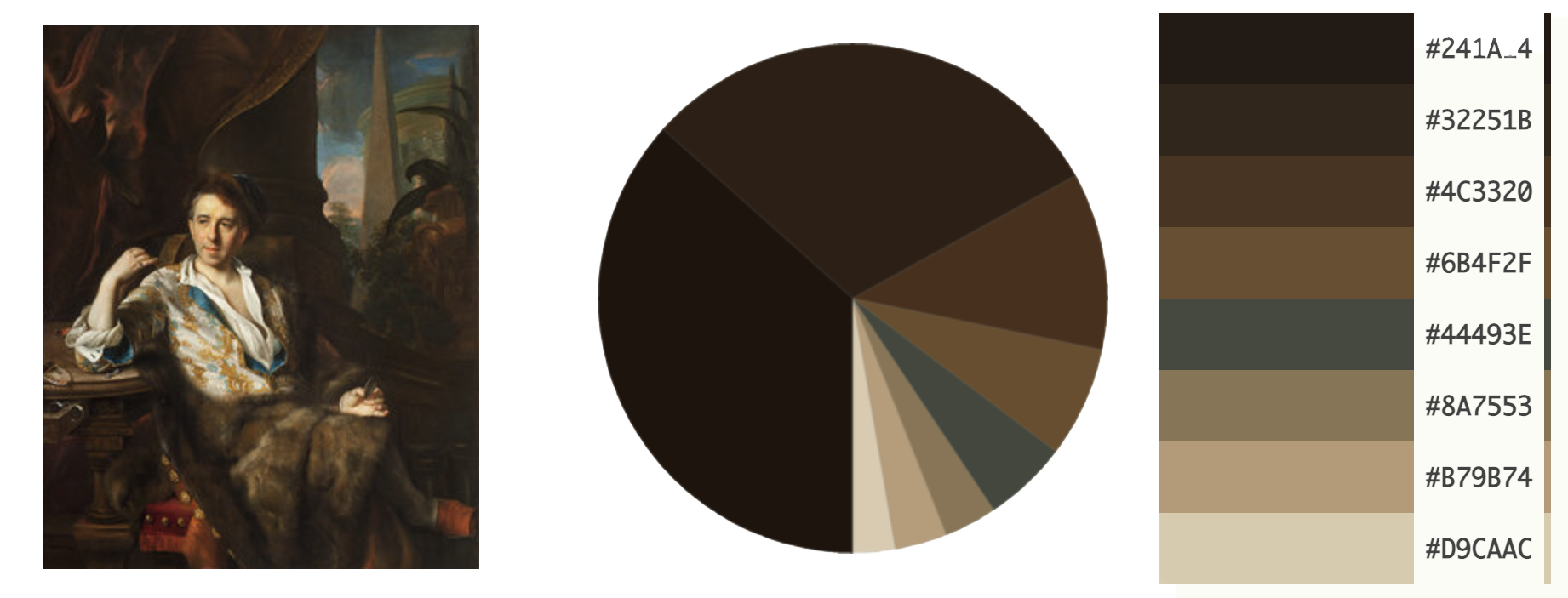}
\caption{\textbf\scriptsize{Dominant color palette of 'Portrait of the Painter Charles Bruni' by Johann Kupetzky.}}
\end{figure}

\begin{figure}[!h]
\centering
\includegraphics[width=1\textwidth]{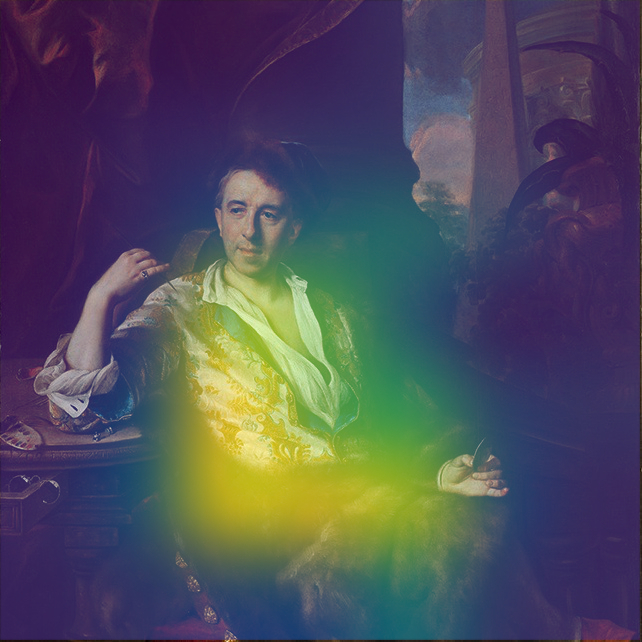}
\caption{\textbf\scriptsize{Grad-CAM class activation visualization from the proposed model.}}
\end{figure}

\section{Conclusions}
This study investigated the design and construction of a CNN-based model for fine art classification using transfer learning and feature extraction utilizing the freely available dataset 'Art Snobs Data2' from the Kaggle repository and evaluated its performance on the example of 'Portrait of the Painter Charles Bruni' by Johann Kupetzky. The classification results demonstrated that the proposed approach provides a computationally efficient way to classify artworks by fine art style without the necessity of defining and fully training new CNN model structures or increasing the size of existing databases. Empirical results based on the confusion matrix demonstrate that CNN has learned relevant features for individual classes, which can be proven by misclassification when trying to classify a work of art across historically related styles.

Although the classifier has achieved interesting results and proposed possible applications, in the future we would like to extend this research to the context of contemporary curatorial practice. While the CNN only predicts the outcome on the basis of visual input (individual shapes, edges, or colors) without any knowledge of the sociocultural context, the human expert interprets the works according to complex relevant features and characteristics that the CNN cannot observe. It is difficult to assess whether it is possible to simply define individual art movements, to evaluate a work using visual interpretation alone, or whether such an interpretation is even relevant to existing curatorial practice; nevertheless, this research has outlined possible pathways for future research.

\bibliographystyle{plainnat}
\bibliography{references}

\begin{thebibliography}{36}
\providecommand{\natexlab}[1]{#1}
\providecommand{\url}[1]{\texttt{#1}}
\expandafter\ifx\csname urlstyle\endcsname\relax
  \providecommand{\doi}[1]{doi: #1}\else
  \providecommand{\doi}{doi: \begingroup \urlstyle{rm}\Url}\fi

\bibitem[Albawi et~al.(2017)Albawi, Mohammed, and
  Al-Zawi]{albawi2017understanding}
Saad Albawi, Tareq~Abed Mohammed, and Saad Al-Zawi.
\newblock Understanding of a convolutional neural network.
\newblock In \emph{2017 international conference on engineering and technology
  (ICET)}, pages 1--6. Ieee, 2017.

\bibitem[Arora and Elgammal(2012)]{arora2012towards}
Ravneet~Singh Arora and Ahmed Elgammal.
\newblock Towards automated classification of fine-art painting style: A
  comparative study.
\newblock In \emph{Proceedings of the 21st International Conference on Pattern
  Recognition (ICPR2012)}, pages 3541--3544. IEEE, 2012.

\bibitem[Aydo{\u{g}}an(2019)]{aydougan2019reflections}
D~Aydo{\u{g}}an.
\newblock Reflections of digitalization on painting.
\newblock \emph{CTC 2019}, 2019.

\bibitem[Cetinic et~al.(2018)Cetinic, Lipic, and Grgic]{cetinic2018fine}
Eva Cetinic, Tomislav Lipic, and Sonja Grgic.
\newblock Fine-tuning convolutional neural networks for fine art
  classification.
\newblock \emph{Expert Systems with Applications}, 114:\penalty0 107--118,
  2018.

\bibitem[Chen et~al.(2016)Chen, Jiang, Li, Jia, and Ghamisi]{chen2016deep}
Yushi Chen, Hanlu Jiang, Chunyang Li, Xiuping Jia, and Pedram Ghamisi.
\newblock Deep feature extraction and classification of hyperspectral images
  based on convolutional neural networks.
\newblock \emph{IEEE Transactions on Geoscience and Remote Sensing},
  54\penalty0 (10):\penalty0 6232--6251, 2016.

\bibitem[Doi(2007)]{doi2007computer}
Kunio Doi.
\newblock Computer-aided diagnosis in medical imaging: historical review,
  current status and future potential.
\newblock \emph{Computerized medical imaging and graphics}, 31\penalty0
  (4-5):\penalty0 198--211, 2007.

\bibitem[Goutte and Gaussier(2005)]{goutte2005probabilistic}
Cyril Goutte and Eric Gaussier.
\newblock A probabilistic interpretation of precision, recall and f-score, with
  implication for evaluation.
\newblock In \emph{European conference on information retrieval}, pages
  345--359. Springer, 2005.

\bibitem[Habsary et~al.(2021)Habsary, Kurniawan, and
  Bulan]{habsary2021digitalization}
Dwiyana Habsary, Agung Kurniawan, and Indra Bulan.
\newblock Digitalization of arts.
\newblock In \emph{Proceedings of the Tenth International Conference on
  Languages and Arts (ICLA 2021)}, pages 246--250. Atlantis Press, 2021.

\bibitem[Hagtvedt and Patrick(2011)]{hagtvedt2011turning}
Henrik Hagtvedt and Vanessa~M Patrick.
\newblock Turning art into mere illustration: Concretizing art renders its
  influence context dependent.
\newblock \emph{Personality and Social Psychology Bulletin}, 37\penalty0
  (12):\penalty0 1624--1632, 2011.

\bibitem[He et~al.(2016)He, Zhang, Ren, and Sun]{he2016deep}
Kaiming He, Xiangyu Zhang, Shaoqing Ren, and Jian Sun.
\newblock Deep residual learning for image recognition.
\newblock In \emph{Proceedings of the IEEE conference on computer vision and
  pattern recognition}, pages 770--778, 2016.

\bibitem[Huang et~al.(2017)Huang, Rathod, Sun, Zhu, Korattikara, Fathi,
  Fischer, Wojna, Song, Guadarrama, et~al.]{huang2017speed}
Jonathan Huang, Vivek Rathod, Chen Sun, Menglong Zhu, Anoop Korattikara,
  Alireza Fathi, Ian Fischer, Zbigniew Wojna, Yang Song, Sergio Guadarrama,
  et~al.
\newblock Speed/accuracy trade-offs for modern convolutional object detectors.
\newblock In \emph{Proceedings of the IEEE conference on computer vision and
  pattern recognition}, pages 7310--7311, 2017.

\bibitem[Joshi et~al.(2019)Joshi, Verma, Saxena, and Paraye]{joshi2019issues}
Soumya Joshi, Dhirendra~Kumar Verma, Gaurav Saxena, and Amit Paraye.
\newblock Issues in training a convolutional neural network model for image
  classification.
\newblock In \emph{International Conference on Advances in Computing and Data
  Sciences}, pages 282--293. Springer, 2019.

\bibitem[Khoronko and Mokina(2021)]{khoronko2021museum}
Lubov Khoronko and Anna Mokina.
\newblock Museum practice in the developing of applied artists’ professional
  competencies in the context of digitalization of education.
\newblock In \emph{E3S Web of Conferences}, volume 273, page 12066. EDP
  Sciences, 2021.

\bibitem[Kim et~al.(2016)Kim, Lee, and Lee]{kim2016accurate}
Jiwon Kim, Jung~Kwon Lee, and Kyoung~Mu Lee.
\newblock Accurate image super-resolution using very deep convolutional
  networks.
\newblock In \emph{Proceedings of the IEEE conference on computer vision and
  pattern recognition}, pages 1646--1654, 2016.

\bibitem[Lindsay(2021)]{lindsay2021convolutional}
Grace~W Lindsay.
\newblock Convolutional neural networks as a model of the visual system: Past,
  present, and future.
\newblock \emph{Journal of cognitive neuroscience}, 33\penalty0 (10):\penalty0
  2017--2031, 2021.

\bibitem[Lu and Weng(2007)]{lu2007survey}
Dengsheng Lu and Qihao Weng.
\newblock A survey of image classification methods and techniques for improving
  classification performance.
\newblock \emph{International journal of Remote sensing}, 28\penalty0
  (5):\penalty0 823--870, 2007.

\bibitem[Martin(2019)]{martin_2019}
Parker Martin.
\newblock Art snobs data2, Nov 2019.
\newblock URL
  \url{https://www.kaggle.com/datasets/parkerzmartin/art-snobs-data2}.

\bibitem[Nash(2008)]{nash2008northern}
Susie Nash.
\newblock \emph{Northern renaissance art}.
\newblock OUP Oxford, 2008.

\bibitem[Nixon and Aguado(2019)]{nixon2019feature}
Mark Nixon and Alberto Aguado.
\newblock \emph{Feature extraction and image processing for computer vision}.
\newblock Academic press, 2019.

\bibitem[O'Shea and Nash(2015)]{o2015introduction}
Keiron O'Shea and Ryan Nash.
\newblock An introduction to convolutional neural networks.
\newblock \emph{arXiv preprint arXiv:1511.08458}, 2015.

\bibitem[Perez and Wang(2017)]{perez2017effectiveness}
Luis Perez and Jason Wang.
\newblock The effectiveness of data augmentation in image classification using
  deep learning.
\newblock \emph{arXiv preprint arXiv:1712.04621}, 2017.

\bibitem[Rodriguez et~al.(2018)Rodriguez, Lech, and
  Pirogova]{rodriguez2018classification}
Catherine~Sandoval Rodriguez, Margaret Lech, and Elena Pirogova.
\newblock Classification of style in fine-art paintings using transfer learning
  and weighted image patches.
\newblock In \emph{2018 12th International Conference on Signal Processing and
  Communication Systems (ICSPCS)}, pages 1--7. IEEE, 2018.

\bibitem[Ruth and Kolehmainen(1974)]{ruth1974classification}
Jan-Erik Ruth and Ky{\"o}sti Kolehmainen.
\newblock Classification of art into style periods; a factor-analytical
  approach.
\newblock \emph{Scandinavian Journal of Psychology}, 15\penalty0 (1):\penalty0
  322--327, 1974.

\bibitem[Saleh and Elgammal(2015)]{saleh2015large}
Babak Saleh and Ahmed Elgammal.
\newblock Large-scale classification of fine-art paintings: Learning the right
  metric on the right feature.
\newblock \emph{arXiv preprint arXiv:1505.00855}, 2015.

\bibitem[Sandoval et~al.(2019)Sandoval, Pirogova, and Lech]{sandoval2019two}
Catherine Sandoval, Elena Pirogova, and Margaret Lech.
\newblock Two-stage deep learning approach to the classification of fine-art
  paintings.
\newblock \emph{IEEE Access}, 7:\penalty0 41770--41781, 2019.

\bibitem[Schwarting et~al.(2018)Schwarting, Alonso-Mora, and
  Rus]{schwarting2018planning}
Wilko Schwarting, Javier Alonso-Mora, and Daniela Rus.
\newblock Planning and decision-making for autonomous vehicles.
\newblock \emph{Annual Review of Control, Robotics, and Autonomous Systems},
  1:\penalty0 187--210, 2018.

\bibitem[Sengupta et~al.(2019)Sengupta, Ye, Wang, Liu, and
  Roy]{sengupta2019going}
Abhronil Sengupta, Yuting Ye, Robert Wang, Chiao Liu, and Kaushik Roy.
\newblock Going deeper in spiking neural networks: Vgg and residual
  architectures.
\newblock \emph{Frontiers in neuroscience}, 13:\penalty0 95, 2019.

\bibitem[Shorten and Khoshgoftaar(2019)]{shorten2019survey}
Connor Shorten and Taghi~M Khoshgoftaar.
\newblock A survey on image data augmentation for deep learning.
\newblock \emph{Journal of big data}, 6\penalty0 (1):\penalty0 1--48, 2019.

\bibitem[Szegedy et~al.(2016)Szegedy, Vanhoucke, Ioffe, Shlens, and
  Wojna]{szegedy2016rethinking}
Christian Szegedy, Vincent Vanhoucke, Sergey Ioffe, Jon Shlens, and Zbigniew
  Wojna.
\newblock Rethinking the inception architecture for computer vision.
\newblock In \emph{Proceedings of the IEEE conference on computer vision and
  pattern recognition}, pages 2818--2826, 2016.

\bibitem[Tan et~al.(2016)Tan, Chan, Aguirre, and Tanaka]{tan2016ceci}
Wei~Ren Tan, Chee~Seng Chan, Hern{\'a}n~E Aguirre, and Kiyoshi Tanaka.
\newblock Ceci n'est pas une pipe: A deep convolutional network for fine-art
  paintings classification.
\newblock In \emph{2016 IEEE international conference on image processing
  (ICIP)}, pages 3703--3707. IEEE, 2016.

\bibitem[Weiss et~al.(2016)Weiss, Khoshgoftaar, and Wang]{weiss2016survey}
Karl Weiss, Taghi~M Khoshgoftaar, and DingDing Wang.
\newblock A survey of transfer learning.
\newblock \emph{Journal of Big data}, 3\penalty0 (1):\penalty0 1--40, 2016.

\bibitem[Wiatowski and B{\"o}lcskei(2017)]{wiatowski2017mathematical}
Thomas Wiatowski and Helmut B{\"o}lcskei.
\newblock A mathematical theory of deep convolutional neural networks for
  feature extraction.
\newblock \emph{IEEE Transactions on Information Theory}, 64\penalty0
  (3):\penalty0 1845--1866, 2017.

\bibitem[Wittkower et~al.(1999)Wittkower, Connors, and
  Montagu]{wittkower1999art}
Rudolf Wittkower, Joseph Connors, and Jennifer Montagu.
\newblock \emph{Art and Architecture in Italy, 1600--1750: Volume 3: Late
  Baroque and Rococo, 1675--1750}, volume~3.
\newblock Yale University Press, 1999.

\bibitem[Yamazaki et~al.(2019)Yamazaki, Kasagi, Tabuchi, Honda, Miwa, Fukumoto,
  Tabaru, Ike, and Nakashima]{yamazaki2019yet}
Masafumi Yamazaki, Akihiko Kasagi, Akihiro Tabuchi, Takumi Honda, Masahiro
  Miwa, Naoto Fukumoto, Tsuguchika Tabaru, Atsushi Ike, and Kohta Nakashima.
\newblock Yet another accelerated sgd: Resnet-50 training on imagenet in 74.7
  seconds.
\newblock \emph{arXiv preprint arXiv:1903.12650}, 2019.

\bibitem[Zhao et~al.(2021)Zhao, Zhou, Qiu, and Jiang]{zhao2021compare}
Wentao Zhao, Dalin Zhou, Xinguo Qiu, and Wei Jiang.
\newblock Compare the performance of the models in art classification.
\newblock \emph{Plos one}, 16\penalty0 (3):\penalty0 e0248414, 2021.

\bibitem[Zhuang et~al.(2020)Zhuang, Qi, Duan, Xi, Zhu, Zhu, Xiong, and
  He]{zhuang2020comprehensive}
Fuzhen Zhuang, Zhiyuan Qi, Keyu Duan, Dongbo Xi, Yongchun Zhu, Hengshu Zhu, Hui
  Xiong, and Qing He.
\newblock A comprehensive survey on transfer learning.
\newblock \emph{Proceedings of the IEEE}, 109\penalty0 (1):\penalty0 43--76,
  2020.

\end{thebibliography}
\addcontentsline{toc}{chapter}{Bibliography} 

\end{document}